\documentstyle[twocolumn,aps,psfig,epsf]{revtex}

\newcommand{\lan}{\langle}
\newcommand{\ran}{\rangle}

\title{Off equilibrium properties of vortex creep in superconductors}
\author{Henrik Jeldtoft Jensen$^{a}$ and Mario Nicodemi$^{a,b}$}
\address{$^a$Department of Mathematics, Imperial College, 180 Queen's Gate, 
London SW7 2BZ, UK\\
$^b$Universit\'a di Napoli ``Federico II'',
Dip. Scienze Fisiche, INFM and INFN, Via Cintia, 80126 Napoli, Italy}

\begin{document}

\maketitle

\begin{abstract}
We study a model for the dynamics of vortices in type II superconductors. 
In particular, we discuss glassy ``off equilibrium'' 
properties and ``aging'' in magnetic creep. 
At low temperatures a crossover point is found, $T_g$, 
where relaxation times seem to diverge \'a la Vogel-Tamman-Fulcher. 
Magnetic creep changes by crossing $T_g$: 
above $T_g$ power law creep is found asymptotically followed by 
stretched exponential saturation; below $T_g$ the creep is logarithmic 
and vortex motion strongly subdiffusive. In this region violation of time 
translation invariance is found along with 
important dynamical scaling properties. 
\end{abstract}


Despite apparent structural differences, 
the {\it dynamics} of vortices in superconductors 
\cite{blatter,Yeshurun}
has many common features with glass formers 
such as supercooled liquids, spin glasses or polymers
\cite{Angell,BCKM}. 
In superconductors even simple quantities such as the sample magnetisation, 
$M$, exhibit strong ``glassy'' behaviours and a variety of 
``history dependent'' effects has been discussed 
\cite{memoryeffects,Papad,our_previous}. 
Common examples are hysteretic magnetisation loops 
which at low temperature exhibit ``memory'' effects: 
the value of $M$ for a given 
applied field, $H$, depends on the sweep rate of $H$, i.e., 
on the history of the sample \cite{memoryeffects}. 
``Aging'' has been also recently observed in magnetic relaxation 
in type II superconductors \cite{Papad}. 
Interestingly, these phenomena are very similar to those 
found in other glassy systems 
such as supercooled liquids. For instance, after a quench, 
``cooling rate'' dependences and ``aging'' effects are usually  
recorded in particle density measurements \cite{Angell}. 

Those similarities are striking when one considers the differences 
which exist at a microscopic level between these systems. 
One is apparent: the interaction ranges of vortices 
in superconductors are typically much larger than those of molecules 
in glass formers. 
In this perspective our aim is to discuss the above 
connections and relate off equilibrium dynamics of vortex matter 
to the dynamics of glass former liquids, where 
relevant progresses have been recently accomplished \cite{BCKM,MP}. 
In all these systems, the glassy features are due to some general mechanisms  
and unrelated to specific material parameters. 
Thus, a statistical mechanics approach, disregarding sample details, 
seems appropriate to describe them. For instance, 
many of the features of glassy behaviours in fragile glasses, 
such as colloidal suspensions, are well described in the context of 
spin glass models \cite{BCKM,MP}. 

Extensive work has been done on the equilibrium or close to 
equilibrium properties of the so-called vortex glass phase proposed 
by M.P.A. Fisher (see ref.s in \cite{blatter}). 
Here we discuss, in particular, the {\em dynamics} of the system 
when {\em far from equilibrium}, a typical situation due to the enormous 
relaxation times found in the low temperatures region. 
We consider a simple tractable model for vortices \cite{our_previous} 
which can be well understood in the context of standard statistical 
mechanics of disordered systems 
\cite{Nattermann}. 
Such a model was already shown to successfully depict 
a unified picture of creep and 
transport phenomena in vortex physics, ranging from magnetisation loops with 
``anomalous'' second peak, logarithmic relaxation at vanishing $T$, 
Bean profiles, 
to history dependent behaviours in vortex flow and I-V characteristics, 
to the reentrant nature of the equilibrium phase diagram \cite{our_previous}. 

As mentioned, here we study in details the system {\em off-equilibrium} 
creep dynamics. The model gives the following scenario for vortex dynamics. 
At low temperatures a crossover point is found, $T_g$, where relaxation
times become longer than the typical observation scale \cite{nota_Tg}. 
Actually, they seem to diverge 
at a lower temperature, $T_c$, where an ``ideal'' thermodynamic glass 
transition can be located (in 2d $T_c$ is numerically zero). 
Magnetic creep changes when $T_g$ is crossed: above $T_g$ power law 
creep is found asymptotically followed by stretched exponentials; 
below $T_g$ the creep is logarithmic and shows strong form of 
``aging'', such as violation of {\em time translation invariance} with 
apparent {\em dynamical scaling properties}. 
Microscopically, $T_g$ corresponds to a drastic change in vortex motion 
which becomes strongly subdiffusive. 
We also outlines the connections of vortex off equilibrium 
dynamics with other glassy systems. 

We study a Restricted Occupancy Model (ROM) \cite{our_previous}, 
a statistical mechanics model which in the limit of zero temperature 
and infinite upper critical field reduces to a cellular automaton 
introduced in \cite{bassler1} to study vortex avalanches. 
The model is a coarse grained version \cite{bassler1,our_previous} 
of a system 
of straight  vortex lines: a set of repulsive particles diffusing in a 
disordered pinning environment.  
The coarse graining length, $l_0$, is taken to be of the order of the natural 
screening length of the problem (which in our case is the London penetration 
length $\lambda$). 
Thus, the coarse graining technique \cite{note0} 
reduces the original interaction potential to an effective short ranged one. 
The price to pay is a loss of information on scales smaller than $l_0$ and 
the necessity to allow multiple occupancy on the lattice sites of the coarse 
grained system. The ROM model is defined by the Hamiltonian 
\cite{our_previous}:
${\cal H}= \frac{1}{2} \sum_{ij} n_i A_{ij} n_j 
-\frac{1}{2} \sum_i A_{ii} n_i - \sum_i A^p_i n_i$,
together with the constraint  $0\leq n_i\leq N_{c2}$  
on the integer occupancy  variable $n_i$ representing the number of vortices 
of the coarse grained cell corresponding to site $i$ (bounded by the upper 
critical field). 

The first term in ${\cal H}$ represents the repulsion between 
particles \cite{blatter}. 
For the above considerations on vortex interactions, 
in the present coarse grained representation (where $l_0\sim \lambda$) 
a finite range potential 
must be considered. For simplicity we choose $A_{ii}=A_0$; $A_{ij}=A_1$ 
if $i$ and $j$ are nearest neighbours; $A_{ij}=0$ for all others couples of sites. 
The second term in ${\cal H}$ concerns the particle self-interaction 
energy and the third one is a pinning potential, with a 
given distribution $P(A^p)$, acting on a fraction $p$ of lattice sites
(below $p=1/2$). 
For simplicity we choose: $P(A^p)=(1-p)\delta(A^p)+p\delta(A^p-A^p_0)$. 
$A_0$ sets the energy scale and we choose 
$A_0=1.0$; $A_0^p=0.3$; $N_{c2}=27$; $\kappa^*\equiv A_1/A_0\in[0,0.3]$
(below we typically discuss data for $\kappa^*=0.28$).
The relation of the parameters of the model to material parameters is 
shown elsewhere \cite{our_previous}, here we only recall that 
$\kappa^*$ is an increasing function of the Ginzburg-Landau $\kappa$ 
and $N_{c2}\propto H_{c2}$. 

Through its surfaces, the system is in contact with 
an external reservoir of ``particles'', 
which schematically corresponds to the applied field present in 
magnetic experiments on superconductors. 
Particles are introduced and escape the system only through the reservoir, 
which, by definition, has a given density $N_{ext}$. 
We performed Monte Carlo Kawasaki (MC) simulations \cite{Binder} on a square 
lattices of linear size $L$ ($L\in\{8,...,128\}$) 
in presence of a thermal bath of temperature $T$. 
Our numerical statistical averages run, according to system size, 
from $128$ to $512$ pinning realizations.


The magnetisation in the system at time $t$ \cite{note3} is defined as:
$M(t)=N_{in}(t)-N_{ext}(t)$, 
where $N_{in}=\sum_i n_i/L^2$ is the overall density inside 
the system and $N_{ext}$ is the applied field. 
As in usual zero field cooled experiments, 
we record the isothermal relaxation of $M(t)$ 
after ramping at a given ``sweep rate'', $\gamma$, the field 
from zero up to a given working value $N_{ext}$. 

The presence, at low temperature, of sweep rate dependent hysteretic cycles, 
slowly relaxing magnetisation, and similar effects \cite{our_previous}, 
indicate that our system, on the observed 
time scales, can be far from equilibrium. Off-equilibrium behaviour is appropriately described by two times correlation functions. 
Thus, along with $M(t)$, we also record the magnetic correlation function ($t>t_w$): 
\begin{equation}
C(t,t_w)
=\langle [M(t)-M(t_w)]^2\rangle ~ .
\label{cor}
\end{equation}
Since the general dynamical features of $C$ and $M$ are very similar, for 
simplicity we mainly discuss $C$ which gives richer information. 

\begin{figure}[ht]
\vspace{-1.6cm}
\centerline{\hspace{-2.5cm}
\psfig{figure=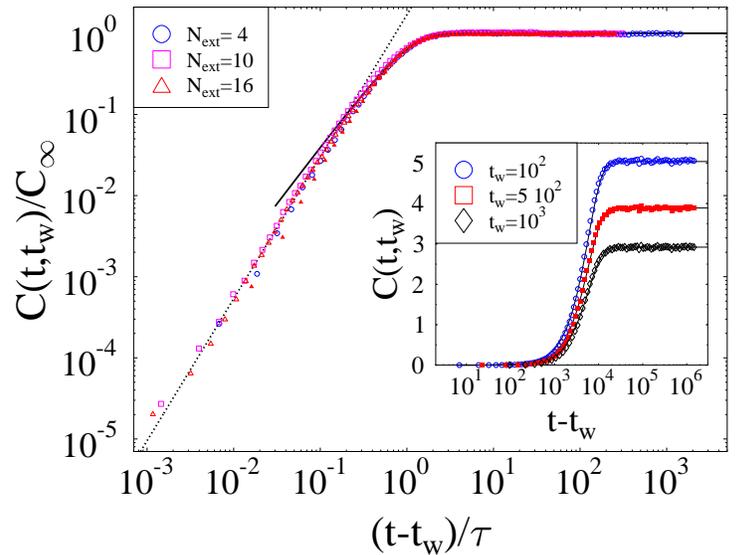,width=8.5cm,angle=-90}}
\vspace{-1.8cm}
\caption{
{\bf Main frame } 
The two time magnetic correlation function $C(t,t_w)$ recorded at $T=1.0$
for the shown values of the external field and waiting times $t_w$. 
All data are collapsed on the same curve when plotted as a function 
of $(t-t_w)/\tau$. 
Here $\tau(N_{ext},T)$ is the characteristic creep time. 
No ``aging'' is present. 
The bold continuous line is a fit to the KWW function of the 
asymptotic region, and the dotted line is a power law fit. 
{\bf Inset } The same data as above for $N_{ext}=16$ as a function of 
$t-t_w$. 
} 
\label{CT1_collapse}
\end{figure}

\begin{figure}[ht]
\vspace{-1.6cm}
\centerline{\hspace{-2.5cm}
\psfig{figure=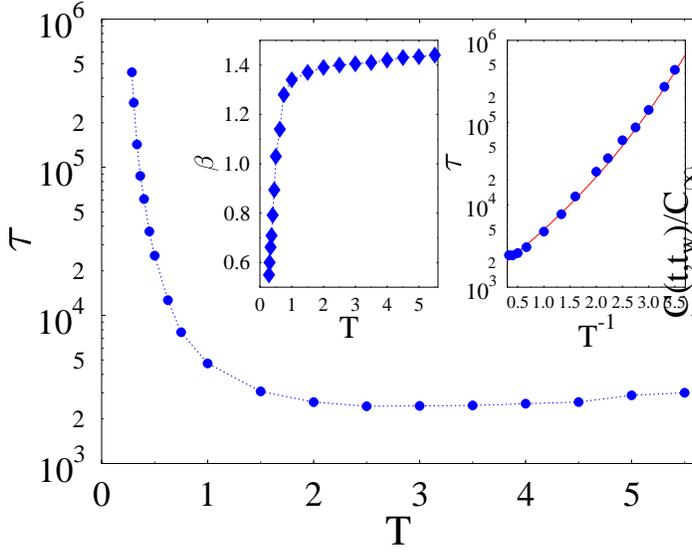,width=8.5cm,angle=-90}}
\vspace{-1.8cm}
\caption{The parameters of the Kohlrausch-Williams-Watts (KWW) 
asymptotic relaxation 
of the magnetisation correlation, $C$, as a function of the temperature $T$, 
recorded at $N_{ext}=10$. 
{\bf Main frame } The equilibration time $\tau$ enormously 
grows by decreasing the temperature $T$. 
Below the crossover temperature $T_g\sim 0.25$, the system 
relaxation times are larger than the observation window. 
{\bf Inset left } The KWW exponent $\beta$ as a function of $T$. 
{\bf Inset right } Close to $T_g$, 
$\tau$ plotted as a function of $1/T$ approximately shows 
a Vogel-Tamman-Fulcher behaviour (see eq.(\ref{VTF})).} 
\label{CT1par_varie_temp}
\end{figure}

At not too low temperatures, for instance at $T=1.0$, the system creep  
is characterised by finite relaxation times, but the dynamics is already 
highly non trivial. 
The two times correlator $C(t,t_w)$ is plotted in Fig.\ref{CT1_collapse} 
after a ramp with $\gamma=10^{-3}$ at $T=1.0$. 
At long times, $C(t,t_w)$ is well fitted by the so called 
Kohlrausch-Williams-Watts (KWW) law
(see Fig.\ref{CT1_collapse}): 
\begin{equation}
C(t,t_w)\simeq C_{\infty}\left\{1-e^{-[(t-t_w)/\tau]^{\beta}}\right\}
\label{strexp_C}
\end{equation}
The KWW decay is observed in the asymptotic relaxation of superconductors 
as well as in glass formers 
(their so called $\alpha$-relaxation \cite{Angell})  
above the glass transition \cite{Angell}. 
The time scale $\tau$ and the Kohlrausch-exponent $\beta$ 
depend on the temperature $T$ (see Fig.\ref{CT1par_varie_temp}) 
and on the overall field $N_{ext}$. 
The pre-asymptotic dynamics (i.e., $t,t_w<<\tau$) is 
also interesting and characterised by various regimes. In particular, for not 
too short times, a power law is observed over several decades 
(see Fig.\ref{CT1_collapse}): 
\begin{equation}
C(t,t_w)\simeq C_0 \left({t-t_w\over \tau}\right)^{a}
\label{power_C}
\end{equation}
The exponent $a$ is almost independent of $N_{ext}$, $a\simeq 1.7$, 
except at very small or high fields.
Notice that the $\tau$ in eq.(\ref{power_C}) is the {\em same} as
in eq.(\ref{strexp_C}), but the exponents $a$ and $\beta$ are 
numerically different (see Fig.\ref{CT1par_varie_temp}). 

Interestingly, the above finding in the model of power laws followed by 
KWW relaxations is typically observed, for not too low $T$, 
in superconductors \cite{Yeshurun}. This behaviour is also typical 
of supercooled liquids, where the power law  regime is called 
the $\beta$-relaxation \cite{Angell}.

At $T=1.0$ or, generally, at not too low temperatures, 
no ``aging'' is seen: 
$C(t,t_w)$ is a function of $t-t_w$. This is clearly shown 
in Fig.\ref{CT1_collapse}, 
where we plot $C(t,t_w)/C_{\infty}$ for several different values 
of $N_{ext}$ as a function of the scaling variable 
$(t-t_w)/\tau$: all the data (for all $t_w$ and $N_{ext}$) 
fall on the same master function (which is more general than the 
above KWW fit). 

The scenario described for $T=1.0$ is found in a broad 
region at low temperatures. However, around $T=0.5$, 
$\beta$ drastically decreases and, at the same time, a steep 
increase of $\tau$ is found (see Fig.\ref{CT1par_varie_temp}). 
For instance, at $N_{ext}=10$ for temperatures below 
$T_g\simeq 0.25$, the characteristic time gets larger  
than our recording window. Thus, below $T_g(N_{ext})$ the system definitely 
loses contact with equilibrium during our observation and, as shown 
in details below, typical glassy phenomena, such as  ``aging'', are observed. 
The crossover temperature $T_g$ is itself a function of $\gamma$. It has 
a physical meaning similar to the so called phenomenological definition of 
the glass transition point in supercooled liquids. Exploiting this analogy 
we will call this temperature the glass temperature despite the fact 
that in glassy systems it is loosely defined \cite{Angell}. The presence of an 
underlying ``ideal'' glass transition point, $T_c(N_{ext})$, is a non trivial 
possibility which, in many cases (as supercooled liquids) is still under debate
\cite{MP,Krauth}. 
$T_c$ is often located by some fit of raw $\tau$ data collected in the 
high $T$ regime \cite{Angell}. 

In our model, in the region where $\tau$ has a steep increase 
a Vogel-Tamman-Fulcher law (VTF) fits the data \cite{note1} 
(see inset of Fig.\ref{CT1par_varie_temp}):
\begin{equation}
\tau=\tau_0 \exp\left({E_0\over T-T_c}\right)
\label{VTF}
\end{equation} 
For example, at $N_{ext}=10$, 
the characteristic time $\tau_0$ is very large, $\tau_0=1.6\cdot 10^3$, 
and the characteristic activation energy, $E_0$, 
is ten times larger than $T_c$: $E_0=1.1$ and $T_c=0.1$, thus 
the above fit is consistent with an Arrhenius low 
(i.e., with $T_c=0$). 
The presence of a strong increase of $\tau$ close to a power law or a 
VTF law is again a mark of the apparent similitude with 
glassy features of supercooled liquids and glasses \cite{Angell}. 

Since below $T_g$ relaxation times are huge, 
one might expect that the motion of the particles essentially stops, 
apart from their vibration inside cages of other vortices. 
Instead, as shown below, 
the off equilibrium dynamics has remarkably rich properties. 
In particular, the properties of the system slowly evolve with time. 
These phenomena are usually summarised by claiming that the systems 
is ``aging''. They occur in many different systems 
ranging from polymers, to supercooled liquids \cite{Angell}, 
spin glasses \cite{BCKM} or granular media \cite{NC_aging}, 
and their origin and apparent universality are still 
an open problem \cite{BCKM,MP,CoNi}.

\begin{figure}[ht]
\vspace{-1.6cm}
\centerline{\hspace{-2.5cm}
\psfig{figure=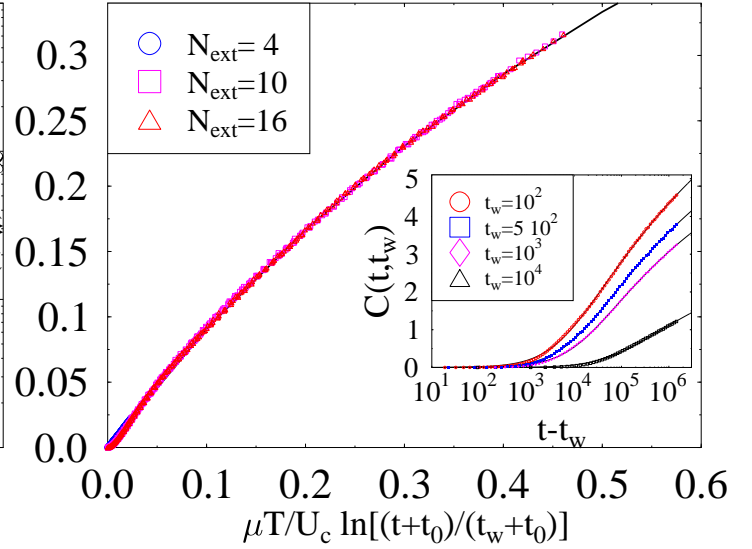,width=8.5cm,angle=-90}}
\vspace{-1.8cm}
\caption{{\bf Inset} 
Logarithmic time relaxation at $T=0.1$ (i.e., below $T_g$) of the two-times 
correlation function, $C(t,t_w)$, recorded at $N_{ext}=16$ (a value close 
to the 2nd peak in magnetic loops). 
{\bf Main Frame} Off equilibrium dynamical scaling. 
Superimposed on the 
same master function, $1-(1+x)^{-1/\mu}$  ($\mu^{-1}\sim 1$), 
are relaxation data of $C(t,t_w)$ recorded for 
$N_{ext}=4,10,16$ for each of the shown $t_w$. 
The asymptotic scaling  is $C(t,t_w)\sim{\cal S}(t/t_w)$.
} 
\label{rel_sca}
\end{figure}

In the inset of Fig.\ref{rel_sca} we show that 
$C(t,t_w)$, at $T=0.1$, exhibits strong ``aging''.  
Notice that $C$ depends on both times $t$ {\em and} $t_w$:
the system evolution is slower the older is its ``age'' $t_w$  
(here, as in eq.(\ref{cor}), $t_w$ is the time elapsed from the sample 
preparation at the working field and temperature). 
Note the contrast with the case at $T=1.0$ of Fig.\ref{CT1_collapse}. 
Such a dynamical ``stiffening'' is typical of glass formers 
\cite{Angell}. It is important to stress 
that the presence of slowly relaxing quantities, such as $M(t)$, 
must not lead to conclude that the system is  
close to equilibrium \cite{BCKM}. 
In the entire low $T$ region ($T<T_g$), after a short initial power law 
behaviour, $C(t,t_w)$ can be well 
fitted by a generalisation of a known interpolation formula, 
often experimentally used \cite{blatter}, which now depends on 
the {\em waiting time}, $t_w$: 
\begin{equation}
C(t,t_w)\simeq C_{\infty}
\left\{1-\left[1+
\frac{\mu T}{U_c}\ln\left(\frac{t+t_0}{t_w+t_0}\right)\right]^{-1/\mu}\right\} 
\end{equation}
We found that to take $\mu\simeq 1$ is consistent with our data. 
In the above fit $U_c/\mu T$ only depends on $N_{ext}$ and, 
interestingly, $t_0$ is approximately a linear function of $t_w$: 
$t_0\propto t_w+t_0^*$, where $t^*_0$ is a constant. 
Very interesting is 
the presence of {\em scaling properties} of purely dynamical origin in the 
off-equilibrium relaxation. 
This is shown in Fig.\ref{rel_sca}, where data for  
different fields, $N_{ext}$, and different waiting times, $t_w$, are rescaled 
on a single master function. The above results imply that 
for times large enough (but smaller than the 
equilibration time), $C(t,t_w)$ is a universal function of the ratio $t/t_w$: 
$C(t,t_w)\sim {\cal S}(t/t_w)$. Such a behaviour 
(called ``simple aging'' \cite{BCKM}) is in agreement with 
a general scenario of off-equilibrium dynamics (see  Ref.\cite{CoNi}) 
and has strong analogies with other glassy systems 
\cite{Angell,BCKM,NC_aging,CoNi}. 
Experimental data on $C(t,t_w)$ do not exist in vortex matter, 
but would be extremely important to shed light on the real nature of dynamical 
phenomena in superconductors. 

Finally, we stress that in vortex physics transitions from low temperature 
logarithmic to higher temperature power law creep are usually experimentally 
found (see references in \cite{Yeshurun}) and give a way to approximately 
locate the position of $T_g$ in real samples.

In the above scenario, microscopic quantities 
concerning the internal rearrangement of the system, 
such as the vortex mean square displacement, $R^2(t)$, can be insightful. 
Since vortices can enter and exit the sample, $R^2(t)$ must 
be properly defined. To this aim, we also made 
a different kind of computer runs 
where after ramping the field from zero to $N_{ext}$ we close the system 
(i.e., remove the reservoir).  
Since now the number of particles is fixed \cite{note2}, their  positions, 
$\vec r_i(t)$, are well defined at each time step and we record: 
$R^2(t)={1\over N}\lan \sum_i({\vec r}_i(t)-{\vec r}_i(0))^2\ran$, 
were $N$ is the total number of present particles. 

\begin{figure}[ht]
\vspace{-1.6cm}
\centerline{\hspace{-2.5cm}
\psfig{figure=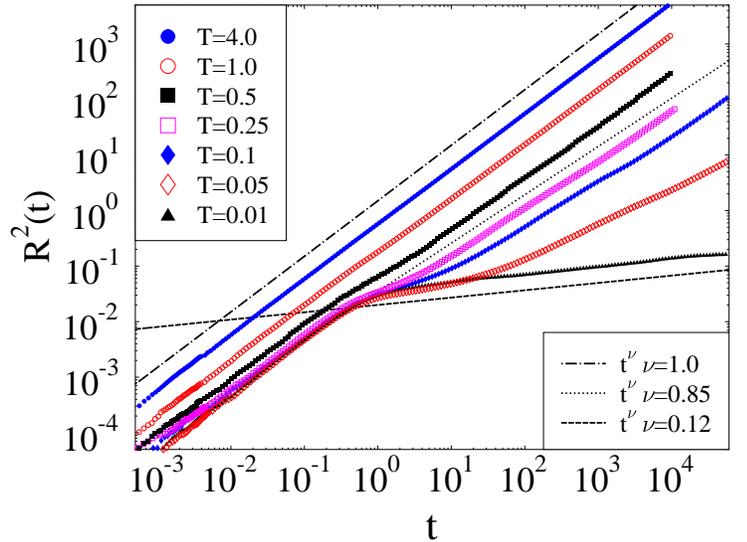,width=8.5cm,angle=-90}}
\vspace{-1.8cm}
\caption{The vortex mean square displacement $R^2(t)$ 
at $N_{ext}=10$ for several temperatures. 
Below $T_g\sim 0.25$, $R^2(t)$ is strongly subdiffusive: $R^2(t)\sim t^{\nu}$ 
with $\nu<1$. Straight lines are guides for the eye. 
} 
\label{r2_fig}
\end{figure}

$R^2(t)$ is plotted in Fig.\ref{r2_fig} for $N_{ext}=10$. At high enough $T$, 
$R^2(t)$ is linear in $t$, but at lower temperatures it shows a 
pronounced bending. Finally, 
below $T_g$, the process at long times becomes subdiffusive: 
\begin{equation}
R^2(t)\sim t^{\nu} 
\end{equation}
with $\nu<< 1$. From this point of view, $T_g$ 
is the location of a sort of structural arrest of the system, where 
particle displacement is dramatically suppressed. 
Interestingly, a very similar scenario has been recorded in real 
superconducting samples: for instance in Ref.\cite{Fuchs} 
it was clearly shown that vortices are definitely mobile 
in the low temperature phase and only ``freeze'' below a 
characteristic field dependent temperature.


Summarising, the schematic coarse grained model we considered  
shows magnetic properties very close to those experimentally found 
in superconductors 
\cite{our_previous}. 
Here, in particular, we have focused on the off equilibrium creep 
dynamics and an interesting scenario emerges.
At low temperatures the system relaxation time $\tau(N_{ext},T)$ grows 
enormously with an ``ideal'' divergence \'a la VTF at 
some $T_c(N_{ext})$ (interestingly, 
molecular dynamics simulations of London-Langevin models 
seem to confirm these results \cite{our_previous,Zimanyi}). 
In fact, below a crossover temperature, $T_g>T_c$, 
vortex displacement becomes sub-diffusive, 
signalling the presence of a structural arrest.  
It is impossible to equilibrate the system on the observation time scale. 
At low temperatures creep is logarithmic and becomes a power law 
above the crossover $T_g(N_{ext})$. 

In typical experiments or computer simulations, magnetic properties are 
measured after ramping the external field at a given rate $\gamma$. 
Whenever $\gamma$ is much larger than the inverse of the characteristic 
relaxation time, $\tau(N_{ext},T)$, the system is naturally driven 
off equilibrium, simply because it is unable to follow the drive, and  
``hysteresis'', ``memory'' effects, along with dependences 
on the sweep rate, occur. In the present framework, 
the strict correspondence 
between the dynamics of vortices and other glass formers 
can be also rationalised. 

Work partially supported by EPSRC, INFM-PRA(HOP)/PCI.

\begin{thebibliography}{0}
\bibitem{nota_Tg} At $T_g$ the systems loses contact with equilibrium, 
but it isn't the ``2nd magnetisation peak'' \cite{our_previous}.
\bibitem{note3} Time is measured in such a way that unity corresponds on  
average to a full lattice sweep.
\bibitem{note1} 
A power law fit is also acceptable: 
$\tau\sim (T-T_c)^{-\gamma}$, with $T_c\simeq 0.21$ and $\gamma\simeq 2.2$.
\bibitem{note2} Creation-annihilation processes are suppressed 
at low $T$ or high fields.
\end{thebibliography}
\end{document}